# Quantum Pseudodots Under the External Vector and Scalar Fields

M. Eshghi[1,*] and S. M. Ikhdair[2,3]

[1] *Department of Physics, Imam Hossein Comprehensive University, Tehran, Iran*
[2] *Department of Physics, Faculty of Science, An-Najah National University, Nablus, Palestine*
[3] *Department of Electrical Engineering, Near East University, Nicosia, Northern Cyprus, Mersin 10, Turkey*

## Abstract

We study the spherical quantum pseudodots in the Schrodinger equation using the pseudo-harmonic plus harmonic oscillator potentials considering the effect of the external electric and magnetic fields. The finite energy levels and the wave functions are calculated. Furthermore, the behavior of the essential thermodynamic quantities such as, the free energy, the mean energy, the entropy, the specific heat, the magnetization, the magnetic susceptibility and the persistent currents are also studied using the characteristic function. Our analytical results are found to be in good agreement with the other works. The numerical results on the energy levels as well as the thermodynamic quantities have also been given.

**Keywords:** Schrödinger equation; harmonic oscillator, pseudo-harmonic potential; thermodynamic quantities; quantum pseudo-dots; characteristic function.

**PACS numbers:** 03.65.Ge; 65.80.-g;

## 1. Introduction

Calculation of the physical quantities in many physical sciences is the essential work we need to perform. As a consequence, the exact solutions of the Schrödinger and Dirac wave equations have become the essential part from the beginning of quantum mechanics [1] and such solutions have also become useful in the fields of the atomic and nuclear physics [2-9].

Currently recent researches on nanometer scale have opened new fields in fundamental sciences of physics, chemistry and engineering such as the optoelectronic devices, high performance laser and detectors [10, 11], where it is called the nanoscience [12]. One of the areas in nanoscience is the class of spherical quantum pseudodots (QPDs). In fact, the spherical QPDs confinement is one of the most appealing explored applications of semiconductor structures when it is doped with shallow donor impurities. Namely, the impurities are used in both transport and optical properties of physics. However, some of the researchers have been extensively studied on the topics like confined donors or acceptors in nanostructures [13–16]. The structure of the QPDs is mainly confined by the quantum potentials and it is also as a result of the recent advances made in the semiconductor which both electrons and valence holes can be confined in all three dimensions [17].

---

[*] ***Email Corresponding Author:*** *eshgi54@gmail.com; m.eshghi@semnan.ac.ir*

One of these interaction potentials used is the pseudo-harmonic potential [18, 19]. The confined pseudo-harmonic potential is generally applied to explain the ro-vibrational states of diatomic molecules and nuclear rotation and vibration. Further, an electron placed in such a potential field is being affected to an external electric field which is equivalent to a pseudo-harmonic oscillator in an external dipole field or a charged pseudo-harmonic oscillator in the presence of a uniform electric field. Such a system has an essential role in quantum chemical applications [20].

On the other hand, confined harmonic oscillator potential can be used to obtain exactly the Schrödinger equation with the pseudo-harmonic and harmonic oscillator potentials and also to find any *l*-state solutions in the view of molecular physics phenomenon. However, in quantum physics, in obtaining the exact solutions of the Schrödinger equation for the molecular potentials can be considered as being one of the main problems [1]. It is well-known that the topic of rotational and molecular vibrational spectroscopy is one of the essential **areas** of molecular physics and it can be counted as one of the main implements for other scientific areas such as environmental sciences [21] and biology [1]. However, the harmonic oscillator could be served as a background to describe the molecular vibrations [22]. However, to improve the theory of molecular vibrations the anharmonic oscillators can be applied to solve exactly the Schrodinger equation and provide more reliable model for diatomic molecules [23].

Here in the present work, one of these interaction potentials used is the pseudo-harmonic oscillator potential [18, 19] plus harmonic oscillator potential taking the form:

$$V(r,z) = V_0 \left( r/r_0 - r_0/r \right)^2 + Kz^2, \tag{1}$$

where the two parameters $V_0$ and $r_0$ stand for the height potential and the zero point, respectively, with $K > 0$ is an oscillator constant.

These confinements lead to the formation of discrete energy levels, drastic change of optical absorption spectra and so forth [24-30].

On the other hand, the study of the thermodynamic properties of low-dimensional semiconductor structure is of a great importance, in particular, determining the behavior of the thermodynamic quantities such as the specific heat, the entropy, the free energy, the mean energy, the magnetization and the persistent current.

Over the past few years, several researchers have studied the thermodynamic properties of various models; see for example, [31-34].

In this work, we solve the Schrödinger equation with the pseudo-harmonic plus harmonic oscillator interaction potential to deal with spherical QPDs being exposed to external electric and magnetic fields. We obtain the finite state energy spectra and their corresponding wave functions. Further, we calculate the essential thermodynamic functions and the thermodynamic quantities by using the finite energy spectrum. Finally we compare our analytical results with those ones obtained by other authors and give a few remarks on the present results.

The organization of the present paper is as follows. In Section 2, we present the solution of the Schrodinger equation with the pseudo-harmonic plus harmonic oscillator potentials exposed to the external electric and magnetic fields for the sake to obtain energy levels and their wave functions. Section 3 is devoted for our results and discussions. Finally, we end with our concluding remarks in Section 4.

## 2. Theory and Calculations

Here we solve the Schrödinger equation with the pseudo-harmonic plus harmonic oscillator interaction potentials so that to calculate the finite bound state energy levels and their corresponding wave functions of the electrons (holes) of spherical QPDs in the presence of external electric and magnetic fields as

$$\left[ \frac{1}{2\mu}\left(\vec{p} - \frac{e}{c}\vec{A}\right)^2 + V(r) + e\vec{\varepsilon}\cdot\vec{z} \right]\psi(r) = E\psi(r), \qquad (2)$$

where $\vec{\varepsilon} = \varepsilon\hat{z}$ is the applied electrostatic field along the $z$-axis, $\mu$ is the effective electronic mass, $A$ is the vector potential which can be found by means of the magnetic field and the energy $E = E_r + E_z$. Let us assume that the vector potential has the simple form: $A = (0, Br/2 + \Phi_{AB}/2\pi r, 0)$ where this potential has recently been used in quantum dots and quantum pseudodots [35, 36]. On substituting Eq. (1) into Eq. (2), the Schrödinger equation reduces into the following forms:

$$\left[ \frac{d^2}{dr^2} + \frac{1}{r}\frac{d}{dr} - \frac{\gamma^2}{r^2} - \omega^2 r^2 + \eta \right] f(r) = 0, \tag{3}$$

and

$$\left[ \frac{d^2}{dz^2} - \left( -\frac{2\mu}{\hbar^2} E_z - \frac{2\mu e.\varepsilon}{\hbar^2} z + \frac{2\mu}{\hbar^2} z^2 \right) \right] G(z) = 0, \tag{4}$$

where the eigenvalues, $E_z$, can be found as

$$E_z = \frac{\hbar}{2}\sqrt{\frac{K}{\mu}}(n_z + 1) - \frac{\hbar^2 e^2 \varepsilon^2}{4K}, \qquad n_z = 1, 2, \dots. \tag{5}$$

Furthermore, we have used the following identifications:

$$\omega^2 = 2\mu V_0/\hbar^2 r_0^2 + e^2 B^2/4\hbar^2 c^2, \qquad \eta = 2\mu(E_n + 2V_0)/\hbar^2 - eB(m+\xi)/\hbar c,$$
$$\gamma^2 = 2\mu V_0 r_0^2/\hbar^2 + (m+\xi)^2, \qquad \xi = \Phi_{AB}/\Phi_0 \text{ with } \Phi_0 = hc/e.. \tag{5a}$$

Now, making the change of variables as $\zeta = \omega r^2$, and hence Eq. (3) can be simply rewritten as

$$\left[ \frac{d^2}{d\zeta^2} + \frac{1}{\zeta}\frac{d}{d\zeta} - \frac{\gamma^2}{4\zeta^2} + \frac{\eta}{4\omega\zeta} - \frac{1}{4} \right] f(\zeta) = 0.. \tag{6}$$

The asymptotic behavior in the solution of Eq. (3) when $r \to 0$ can be determined via the centrifugal term whereas the asymptotic behavior of our solution at $r \to \infty$ can be determined by the oscillating terms. That is, the radial wave function $f(r)$ needs to be finite, using the boundary conditions $f(0) \to 0$ and $f(\infty) \to 0$. Therefore, to make the solution satisfying the above conditions, we are supposed to cast the solution of the wave function $f(\zeta)$ to be $f(\zeta) = \exp(-\zeta/2)\zeta^{|\gamma|/2} F(\zeta)$. Upon substituting this wave function into Eq. (6), we obtain the confluent hypergeometric differential equation,

$$\left[ \zeta \frac{d^2}{d\zeta^2} + (|\gamma| + 1 - \zeta)\frac{d}{d\zeta} - \left( \frac{|\gamma|}{2} + \frac{1}{2} - \frac{\eta}{4\omega} \right) \right] F(\zeta) = 0. \tag{7}$$

The above equation has the solution: $F(\zeta) = (|\gamma|/2 + 1/2 - \eta/4\omega, |\gamma| + 1; \zeta)$. When $\zeta \to \infty$, then $F(\zeta)$ is required to become zero. We further need the confluent hypergeometric series with $F(\zeta)$ to be finite. Notice that when the independent term of Eq. (7) is zero or negative, this requirement is been verified. Therefore, the quantum condition of polynomial confluent function requires that $-n = |\gamma|/2 + 1/2 - \eta/4\omega$. Plugging in the values of parameters $\gamma$, $\omega$ and $\eta$, we can finally obtain the energy spectrum as follows:

$$E_r = \frac{\hbar \omega_c}{2}(m+\xi) - 2V_0 + \sqrt{\hbar^2 \omega_c^2 + \frac{8V_0 \hbar^2}{r_0^2 \mu}} \left[ n + \frac{1}{2} + \frac{1}{2}\sqrt{(m+\xi)^2 + \frac{2\mu V_0 r_0^2}{\hbar^2}} \right], \quad n = 0,1,2,... \quad (8)$$

where $\omega_c = eB/\mu c$ is the cyclotron frequency and $m$ is the projection of the angular momentum on the $z$ axis. If we replace $\gamma \to \beta$ in Eq. (5a), namely, $\gamma^2 = 2\mu V_0 r_0^2/\hbar^2 + (m+\xi)^2$, $\xi \to \alpha$ in Eq. (5a), namely, $\xi = \Phi_{AB}/\Phi_0$ with $\Phi_0 = hc/e$., and $\mu \to m^*$, we have

$$E_r = \hbar \left( \omega_c^2 + \frac{8V_0}{r_0^2 m^*} \right)^{\frac{1}{2}} \left( n + \frac{\beta+1}{2} \right) + \hbar \omega_c \frac{m+\alpha}{2} - 2V_0. \quad (8a)$$

Therefore, Eq. (8a), is same as Eq. (7) of Ref. [19] by Çetin who has calculated the energy states and wave function for an electron confined by a pseudo-harmonic potential both including harmonic dot and antidot potentials under the magnetic and AB flux fields.

Here, we only apply the electrostatic field along the *z*-axis and investigate the thermal properties of spherical QPDs as a complementary study to Çetin work.

Now, under the following two special cases of interest: At first if we put $\Phi_{AB} = 0$ into Eq. (8), then it turns to become Eq. (11) of Refs. [37, 38]. In these works, the authors have calculated the light interband absorption coefficient and the threshold frequency in quantum pseudodot system under the influence of an external magnetic field as when $B(r) = Br/2$. Secondly, when we put both fields $\Phi_{AB} = 0$ and $B = 0$, then Eq. (8) turns into Eq. (18) of Ref. [17].

It is worth noting that in the absence the applied electrostatic field along the *z*-axis, the Eq. (8) in our solution for the finite energy levels are identical to the ones already been found in Eq. (25) of Ref. [35]. It is worthwhile to remark that Ikhdair and Hamzavi have also calculated the interband light absorption coefficient in the quantum pseudodot system with the magnetic field by employing the Nikoforov-Uvarov method [35].

After making use of Eqs. (5) and (8), the total energy levels can be obtained as

$$E_{nm} = n\hbar + a, \quad n = 1, 2, \ldots, \qquad (9a)$$

with the quantum number $n = \sqrt{\left(\dfrac{eB}{\mu c}\right)^2 + \dfrac{8V_0}{r_0^2 \mu} n_r} + \dfrac{1}{2}\sqrt{\dfrac{K}{\mu}} n_z$ and

$$a = \frac{1}{2}\sqrt{\hbar^2\left(\frac{eB}{\mu c}\right)^2 + \frac{8V_0 \hbar^2}{r_0^2 \mu}\left[1 + \sqrt{\left(m + \frac{e\Phi_{AB}}{hc}\right)^2 + \frac{2\mu V_0 r_0^2}{\hbar^2}}\right]}$$
$$+ \frac{\hbar}{2}\frac{eB}{\mu c}\left(m + \frac{e\Phi_{AB}}{hc}\right) - 2V_0 - \frac{\hbar^2 e^2 \varepsilon^2}{4K} + \frac{\hbar}{2}\sqrt{\frac{K}{\mu}}. \qquad (9b)$$

Here the formula (9b) is showing a relationship between parameters $r_0$, $V_0$ and $K$ in the present potential model. Therefore, the solutions of Eq. (9b) are valid for the potential parameters satisfying the restriction given in Eq. (9a). However, the relation between the potential parameters (9b) depends on the azimuthal quantum number $m$ which means that the potential has to be different for various quantum numbers.

Further, the pseudo-harmonic potential plus oscillator potential has a major influence on the energy levels. In the absence of such interaction, the energy levels are obtained as below

$$E_{nm} = \frac{e\hbar B}{2\mu c}\left[2n + 1 + \left|m + \frac{e\Phi_{AB}}{hc}\right| + \frac{1}{2}\left(m + \frac{e\Phi_{AB}}{hc}\right)\right], \quad n = 1, 2, \ldots. \qquad (9c)$$

In Figures 1 to 4, we show the variation of the bound state energies for different parameters. We use a selected parameter values $c = e = r_0 = \hbar = K = \mu = n_r = n_z = 1$, $V_0 = 5$ in plotting these Figures.

For example, in Fig. 1, we plot the pseudodot energy states versus the magnetic field taking $\Phi_{AB} = 5, 10, 15, 20$ with $\varepsilon = 5$ and $m = 1$. In this Figure, it is seen that the energy is changing with the magnetic field. Notice that for a fixed value of the magnetic field, the energy increases with the increasing flux field strength.

In Fig. 2, we draw the pseudodot energy spectrum versus the magnetic field when choosing the radial quantum number values taking $\Phi_{AB} = 5$, $\varepsilon = 5$ and $m = 1$. In this Figure, we see that the pseudodot energy spectrum is increasing with the increasing of the magnetic field. Also, for a specific value of the magnetic field, it is shown the linear variation of the energy with the radial quantum numbers.

In Fig. 3, we draw the pseudodot energy spectrum versus the electric field when choosing the various azimuthal quantum number values with $\Phi_{AB} = 5$ and

$B = 2$. Notice that for a fixed value of the electric field, the energy increases with the increasing values of the azimuthal quantum number.

Fig. 4 shows the pseudodot energy states versus the magnetic field which is increases linearly with the increasing values of the magnetic field for the various values of the azimuthal quantum number.

Now, in working out the thermal properties of spherical QPDs, let us start by defining the fundamental object in statistical physics, that is, the canonical partition function $Z$. Using the energy spectrum of the electrons(holes) of spherical QPDs, Eq. (9), we have

$$\omega_n = \frac{\Omega}{2}(2n + \Xi), \tag{10}$$

where $\omega_n = E_{nm}/\hbar$, $\Omega = 1$ and $\Xi = 2a/\hbar$.

Using the characteristic function ($X = \ln Z$) as below:

$$X = -\sum_{n=1}^{\infty} \ln[1 - \exp(-\beta\omega_{nm})] = -\sum_{n=1}^{\infty} \ln[1 - \exp(-2\pi\delta(2n + \Xi))], \tag{11}$$

where $\delta = \Omega\beta/4\pi$ and $\beta = 1/T$ with supposing $k_B = 1$.

In fact, in the canonical ensemble the thermodynamics of a system such as idea gas of the electrons(holes) in a pseudodot at temperature, $T$, is found from its partition function, $Z = \sum_E \exp(-\beta E)$, with $\beta = 1/K_B T$ where $K_B$ and $E$ denotes the Boltzmann constant and the energy eigenvalues of the system, respectively [35]. Now, an energy value $E$ can be expressed in terms of the single-particle energies, $\varepsilon$; for instance, $E = \sum_k n_k \varepsilon$, where $n_k$ is the number of particles in the single-particle energy state $k$. Therefore, we can write the partition function of gas system as $Z = 1 \Big/ \prod_{n=1}^{\infty}(1 - e^{-\beta\varepsilon_n})$.

Now, the logarithm of $Z$ is known as the characteristic function and denoted by $G = \ln Z$. In fact, a characteristic function is simply the Fourier transform, in probabilistic language. Defining the characteristic function of a random variable $\tilde{X}$ as follows [40, 41]:

$$\tilde{X} = \int_{-\infty}^{\infty} e^{i\omega x} dx, \tag{12}$$

Also, the characteristic function mentions to a particular relation between the partition function of an ensemble in statistical physics. Now, we suppose that $Z$ is the partition function, then, it satisfies $Z = \exp(\pm \beta Q)$ where $Q$ is a thermodynamic quantity. Here,

$Q$ is called as the characteristic function of the ensemble corresponding to $P$. In the micro-canonical ensemble, the patition, $\Omega(U,V,N)$ is as $\Omega(U,V,N)=\exp(\beta TS)$, the canonical ensemble, the partition function is written as $Z(T,V,N)=\exp(-\beta A)$, the partition function is written for the grand canonical ensemble as $\Xi(T,V,\mu)=\exp(-\beta\Phi)$ and the isothermal-isbaric ensemble, the partition function is written as $\Delta(N,T,P)=\exp(-\beta\tilde{G})$, with its characteristic function as $TS$, $A$, $\Phi$ and $\tilde{G}$, respectively. In this area, a list of some common distributions functions and the corresponding characteristic have been given in Ref. [42]. Also, the work statistics for single non-relativistic particle have been determined by Yi and Talkner [43], and $N$-particles by evaluating the characteristic function with the help of the relation into Refs. [44, 45].

At this stage, we can obtain the following relation after expanding the logarithm in Eq. (11) as below:

$$\partial X/\partial\delta = -2\pi\sum_{k=1}^{\infty}e^{-2\pi k\delta(2n+\Xi)}\sum_{n=0}^{\infty}(2n+\Xi), \tag{13}$$

and further using the formula $e^{-x}=(1/2\pi i)\int_C dsx^{-s}\Gamma(s)$ with $x=2\pi k\delta(2n+\Xi)$ in Ref. [46,47], the derivative of X is obtained as follows:

$$\frac{\partial X}{\partial\delta} = -\frac{1}{i}\int_C ds(2\pi\delta)^{-s}\Gamma(s)\zeta(s)\sum_n(2n+\Xi)^{1-s}. \tag{14}$$

Now, Eq. (14) can be written in terms of the Euler, Riemann and Riemann's generalized functions as below:

$$\frac{\partial X}{\partial\delta} = -\frac{1}{i}\int_C ds(2\pi\delta)^{-s}\Gamma(s)\zeta(s)2^{1-s}\zeta\left[s-1,\frac{\Xi}{2}\right]. \tag{15}$$

In eq. (15), we expand $\zeta\left[s-1,\frac{\Xi}{2}\right]$ up to the third order term in $(\Xi-1)$, then, by substituting expanding into Eq. (15), we have

$$\frac{\partial X}{\partial\delta} = -\frac{\pi}{94\delta^2}\left[\frac{1}{4}-(\pi^2-8)(\Xi-1)+(7\zeta[3]-8)(\Xi-1)^2\right]-\frac{\pi}{12}[3\Xi(\Xi+2)+2]+\frac{\Xi+1}{2\delta}, \tag{16}$$

It is worth to mention that we have used the residues for the poles $s-0,1,2$ of Eq. (16). At the end, in the first-order approximation in $1/2-a/\hbar$, we can write the new characteristic function of Eq. (16) as below:

$$X = \left(\frac{1}{2} - \frac{a}{\hbar}\right)\left[\ln\left(\frac{4\pi}{\beta}\right) + \frac{\beta}{2} - \left(2 - \frac{\pi^2}{4}\right)\frac{\pi^2}{3\beta}\right] - \ln\left(\frac{4\pi}{\beta}\right) - \frac{11\beta}{48} + \frac{\pi^2}{6\beta}. \tag{17}$$

Upon using the new characteristic function, the mean energy is calculated as:

$$U = T^2 \frac{\partial X}{\partial T} = \left(\frac{1}{2} - \frac{a}{\hbar}\right)\left[T - \frac{1}{2} - \left(\frac{\pi^2}{4} - 2\right)\pi^2 T^2\right] - T + \frac{11}{48} - \frac{\pi^2 T^2}{6}. \tag{18}$$

The specific heat can be obtained as ($C_V = -\partial U/\partial T$)

$$C_V = \left(\frac{1}{2} - \frac{a}{\hbar}\right)\left[-3 + \frac{2}{T} + \frac{2\pi^2}{3}\left(\frac{\pi^2}{4} - 2\right)T\right] + 1 + \frac{5}{12T} - \frac{\pi^2}{3}T. \tag{19}$$

Further, we can calculate the free energy ($F = -\ln(Z)/\beta$) as below:

$$F = \left(\frac{1}{2} - \frac{a}{\hbar}\right)\left[-\frac{2}{T^2} + \frac{2\pi^2}{3}\left(2 - \frac{\pi^2}{4}\right)T\right] - \frac{5}{12T^2} + \frac{\pi^2}{3}. \tag{20}$$

The entropy is defined as $(S = -\partial F/\partial T)$ and yields

$$S = 4\left(\frac{1}{2} - \frac{a}{\hbar}\right)\frac{1}{T^3} - \frac{5}{6T^3}. \tag{21}$$

For reminder, this physical quantity has been applied in a wide variety of fields and plays a vital role in thermodynamics. Moreover, it is central to the Second law of thermodynamics and helps measure the amount of order and disorder and/or chaos as well. It can be defined and measured in many other fields than the thermodynamics.

Now, the persistent current ($I = -\partial F/\partial \Phi$) [48] can be obtained as below:

$$I = \left[\frac{2}{T^2} - \frac{2\pi^2}{3}\left(2 - \frac{\pi^2}{4}\right)\right]$$

$$\times \left[\sqrt{\frac{\frac{\hbar^2 e^2 B^2}{\mu^2 c^2} + \frac{8V_0 \hbar^2}{r_0^2 \mu}}{\left(m + \frac{e\Phi_{AB}}{hc}\right)^2 + \frac{2\mu V_0 r_0^2}{\hbar^2}}} \frac{\left(m + \frac{e\Phi_{AB}}{hc}\right)\pi e}{\hbar^2 c} + \frac{e^2 B\pi}{\hbar c}\right]. \tag{22}$$

The magnetization ($M = -\partial F/\partial B$) [39, 48] of the present system can be obtained as follows:

$$M = \left[\frac{2}{T^2} - \frac{2\pi^2}{3}\left(2 - \frac{\pi^2}{4}\right)\right]$$

$$\times \left[\frac{\hbar e^2 B\left(1 + \sqrt{\left(m + \frac{e\Phi_{AB}}{hc}\right)^2 + \frac{2\mu V_0 r_0^2}{\hbar^2}}\right)}{2\mu^2 c^2 \sqrt{\frac{\hbar^2 e^2 B^2}{\mu^2 c^2} + \frac{8V_0 \hbar^2}{r_0^2 \mu}}} + \frac{e\left(m + \frac{e\Phi_{AB}}{hc}\right)}{2\hbar c}\right]. \quad (23)$$

Finally, the magnetic susceptibility [49-51] is calculated as

$$\chi = \left[\frac{2}{T^2} - \frac{2\pi^2}{3}\left(2 - \frac{\pi^2}{4}\right)\right]\left[\frac{3\hbar^3 e^4 B\left(1 + \sqrt{\left(m + \frac{e\Phi_{AB}}{hc}\right)^2 + \frac{2\mu V_0 r_0^2}{\hbar^2}}\right)}{2\mu^4 c^4 \left(\frac{\hbar^2 e^2 B^2}{\mu^2 c^2} + \frac{8V_0 \hbar^2}{r_0^2 \mu}\right)^{\frac{3}{2}}}\right.$$

$$\left.+ \frac{3\hbar^5 e^6 B^3\left(1 + \sqrt{\left(m + \frac{e\Phi_{AB}}{hc}\right)^2 + \frac{2\mu V_0 r_0^2}{\hbar^2}}\right)}{2\mu^6 c^6 \left(\frac{\hbar^2 e^2 B^2}{\mu^2 c^2} + \frac{8V_0 \hbar^2}{r_0^2 \mu}\right)^{\frac{5}{2}}}\right]. \quad (24)$$

## 3. Results and Discussion

Here, we present the results of our study. We take parameters values as $c = e = r_0 = \hbar = K = \mu = 1$, $V_0 = 5$ while plotting the Figures. Therefore, we plot Figs. 5 to 8 using Eqs. (18) to (21), respectively. In these Figs. the thermodynamic quantities such as mean energy, specific heat, free energy and entropy quantities are changing with increasing values of temperature *T*. For example, in Fig. 5, this changing is as deceasing, but in Figs. 6 to 8, theses changings are as increasing for different azimuthal quantum numbers *m*.

On the other hand, Figs. 9 to 14 show that the mean energy, specific heat, persistent current, magnetization, magnetic susceptibility and entropy quantities are changing with increasing the pseudodot size for several values of the magnetic field.

It is interesting that this changing is as decreasing in case of Figs. 10, 11 and 14, but this changing is as increasing in case of Figs. 9, 12 and 13. Further, in Fig. 9, it reaches a maximum as slowly at around some *B* for $B \leq 8$, and in Fig. 12, it increase as exponentially. In Fig. 13, it reaches a maximum as Gaussian form at around some

$B$ for $1.5 < B \leq 4$, then, it decrease as exponentially. Also, in Fig. 10, it reaches a minimum at around some $B$ for $B \geq 2$ and in Fig. 11, it decrease as exponentially.

In Figures. 5 to 14, it is seen that the large influence of both the azimuthal quantum number and magnetic field are more apparent, respectively.

We have noticed from Fig. 9 that as $B$ increasing further, the mean energy begins to decrease linearly as well, but in Fig. 10, the specific heat begins to increase linearly as well.

Figure 15 shows that the free energy decreases with increasing the value of the two fields for a few several of the pseudodot size, respectively.

Notice that for a fixed value of the magnetic and AB flux fields, the free energy decreases when the pseudodot size is increasing. Namely, it is obvious to state that the influence of the pseudodot size is more apparent.

Figures 16 to 18 show that the persistent current, magnetization and magnetic susceptibility increase with increasing value of the AB flux field for several of the different azimuthal quantum number. It is seen that the large influence of the azimuthal quantum number is more apparent for magnetic susceptibility.

## 4. Concluding Remarks

We solved the non-relativistic equation with the pseudo-harmonic plus harmonic oscillator potentials under the influence of the magnetic and AB flux fields to study the spherical QPDs. Our results in Eq. (8a), is found to be same as Eq. (7) of Ref. [19].

We calculated the bound states energies and the corresponding wave functions. The finite bound state energies are used to obtain the partition function and then to obtain the main thermodynamic quantities for pseudodot systems. Our results are compared with the results obtained by other authors and found to be in good agreement.

It is worthy to remark that the magnetic susceptibility function reaches its maximum value at the pseudodot size of $r_0 \simeq 4\ nm$ when the magnetic field $B = 2\ T$. However, when the magnetic field strength increases this maximum value will decrease as shown in Figure 13. On the other hand, the maximum of the entropy function curve decreases with the increasing of the pseudodot size $r_0$ for different magnetic field values as shown in Figure 14.

Finally, our results of the energy states are plotted versus the various parameters of this model in Figures 1 to 18.

**Acknowledgements:**

The authors would like to thank the kind referees for the positive suggestions which have greatly improved the present text.

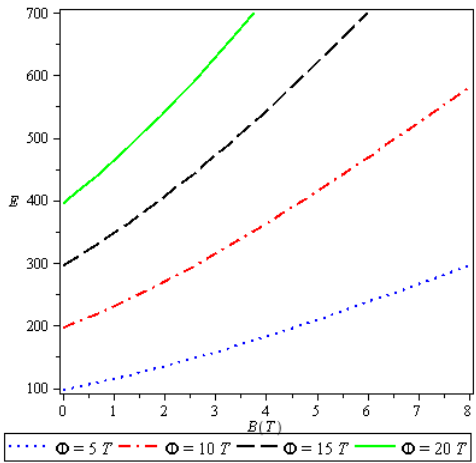

**Figure 1:** The variation of the bound states energies versus the magnetic field for various Aharonov-Bohm flux fields

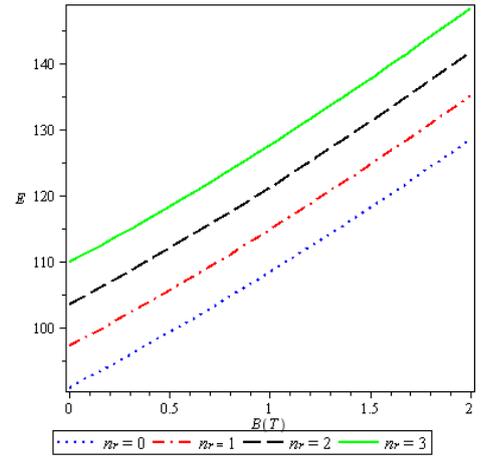

**Figure 2:** The variation of the bound states energies versus the magnetic field for various $n_r$

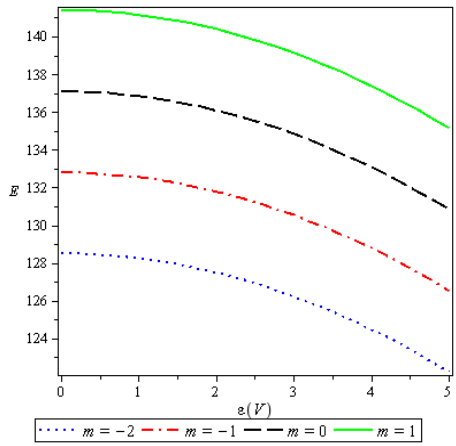

**Figure 3:** The variation of the bound states energies versus the electric field for various $m$

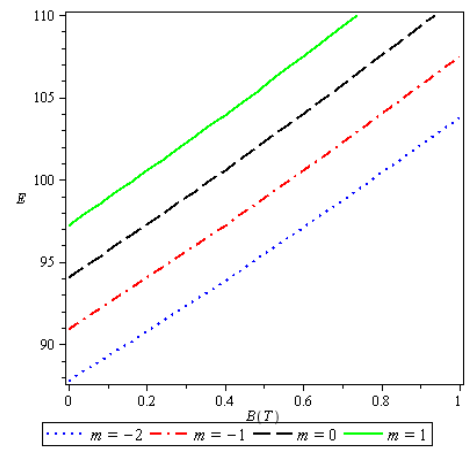

**Figure 4:** The variation of the bound states energies versus the magnetic field for various $m$

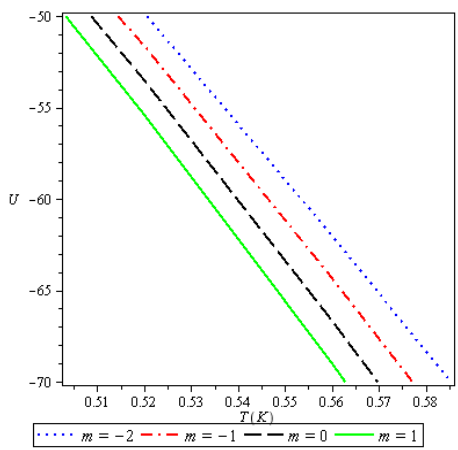

**Figure 5:** The variation of the mean energy function versus the temperature for various $m$

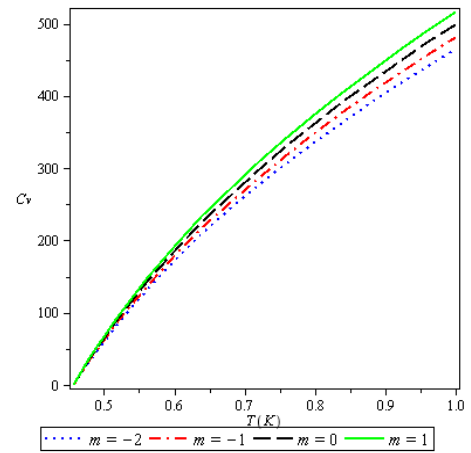

**Figure 6:** The variation of the specific heat function versus the temperature for various $m$

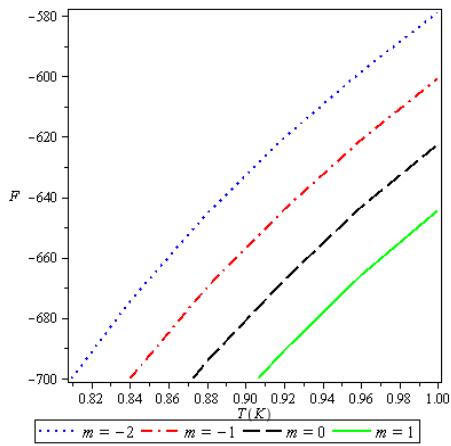

**Figure 7:** The variation of the free energy function versus the temperature for various *m*

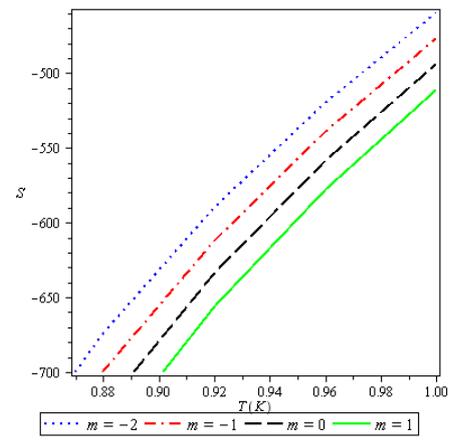

**Figure 8:** The variation of the Entropy function versus the temperature for various *m*

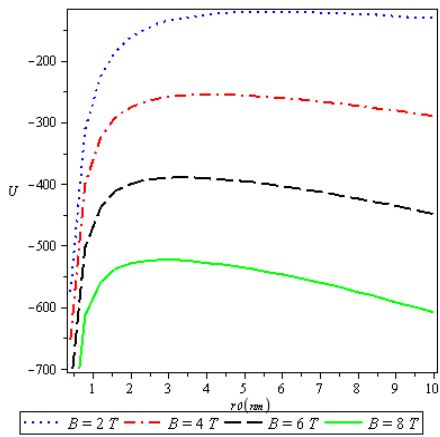

**Figure 9:** The variation of the mean energy function versus the pseudodot size for various magnetic fields

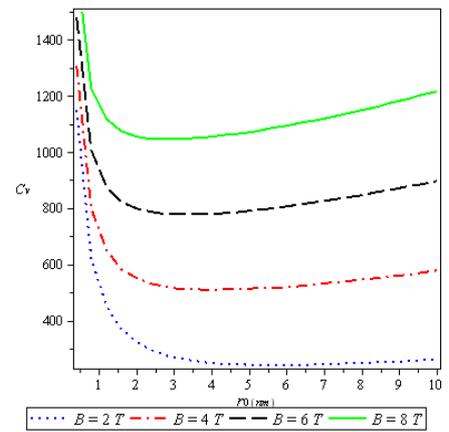

**Figure 10:** The variation of the specific heat function versus the pseudodot size for various *B*

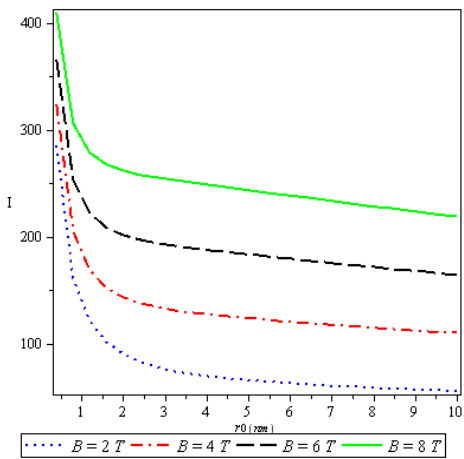

**Figure 11:** The variation of the persistent current function versus the pseudodot size for various *B*

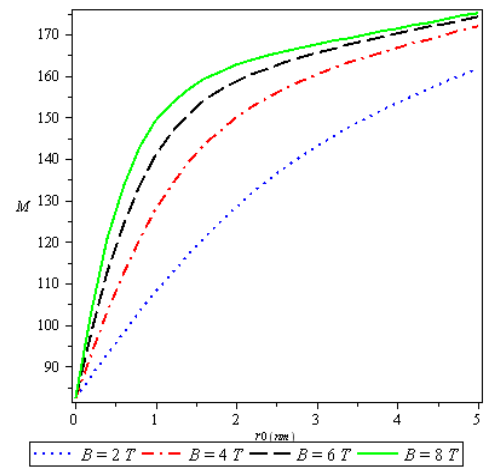

**Figure 12:** The variation of the magnetization function versus the pseudodot size t for various *B*

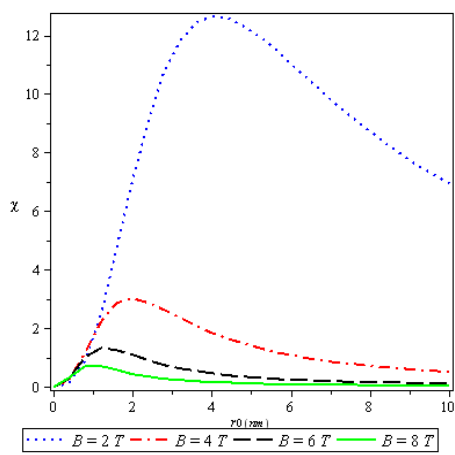

**Figure 13:** The variation of the magnetic susceptibility function versus the pseudodot size for various $B$

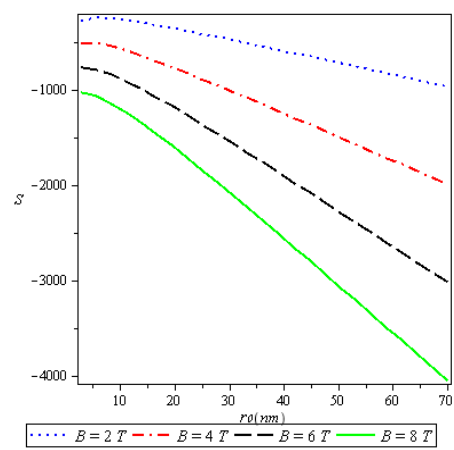

**Figure 14:** The variation of entropy function versus the pseudodot size with the various magnetic fields

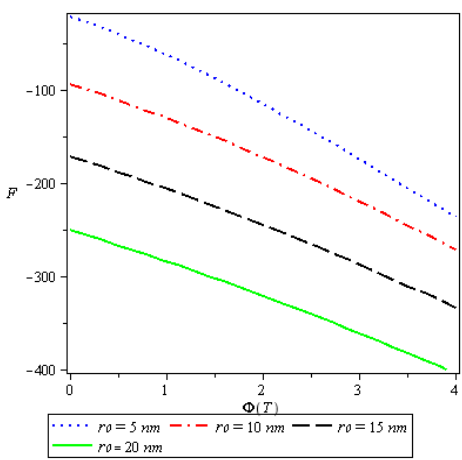

**Figure 15:** The variation of free energy function versus Aharonov-Bohm flux field with the various pseudodot size

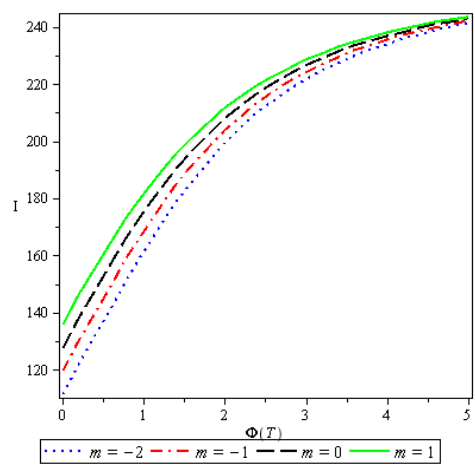

**Figure 16:** The variation of the persistent current function versus the AB flux field for various $m$ with $r_0 = 5$.

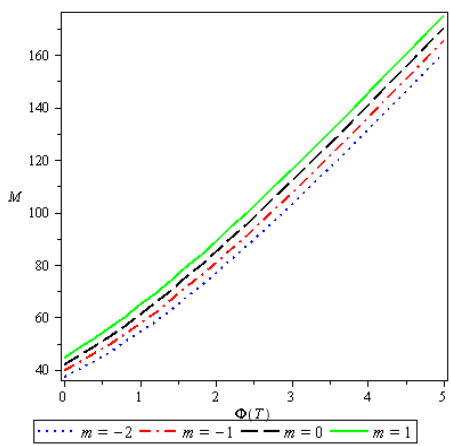

**Figure 17:** The variation of magnetization function versus the Aharonov-Bohm flux field for various $m$ with $r_0 = 5$.

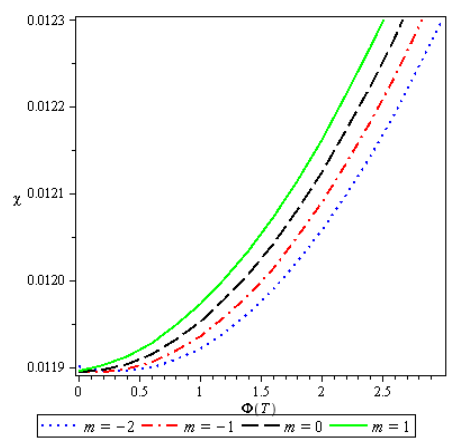

**Figure 18:** The variation of magnetic susceptibility function versus the Aharonov-Bohm flux field for various $m$ with $r_0 = 20$.